\documentstyle[preprint,aps,psfig]{revtex}

\def\be{\begin{equation}}
\def\ee{\end{equation}}
\def\bkk#1{<\kern-0.1167em \kern-0.1167em >} 
\def\etal{{\it et al. }}
\renewcommand{\vec}[1]{\mbox{\boldmath $#1$}}

\tightenlines
\begin{document}
\title{Mean field theory for global binding systematics}
\author{G.F. Bertsch and K. Hagino}
\address{Institute for Nuclear Theory, Department of Physics, 
\\ University of Washington, 
Box 351550, 
Seattle, WA 98195, USA
}
\maketitle

\begin{abstract}

We review some possible improvements of mean field theory for 
application to nuclear binding systematics.  Up to now,
microscopic theory has been less successful than models starting
from the liquid drop in describing accurately the global binding
systematics.  We believe that there are good prospects
to develop a better global theory, using modern forms of 
energy density functionals and treating correlation energies systematically
by the RPA.

\end{abstract}

\section{Introduction}

An important goal of nuclear theory is to predict nuclear binding 
energies. While mean field theory offers 
the most fundamental basis to understand
nuclear structure, paradoxically 
it has not been as successful than other approaches 
in making a global fit to nuclear binding energies.  The most accurate
theory of nuclear binding systematics \cite{mo95} starts from the 
liquid drop model, and treats shell effects perturbatively.  It fits
the binding energies with an RMS deviation of
0.67 MeV, taking 15 free parameters and a similar number of fixed
parameters to achieve the fit.
No such systematic study with so many parameters has been attempted in
a purely microscopic approach, but there are a number of partial studies
beginning with the pioneering work of Vautherin and Brink\cite{VB72} using
an energy functional based on the Skyrme interaction.  Noteworthy recent
papers using Skyrme interactions are by Patyk \etal \cite{pa99} 
and by Brown \cite{br98}.  Patyk \etal find that Skyrme interactions
taken from the literature give RMS errors greater than 2 MeV.
This level is also found for the Gogny interaction, which unlike the
Skyrme is finite range.  Brown recently made a new Skyrme fit 
to closed shell nuclei, including radii and spectroscopic properties in the 
fitting \cite{br98}.
He found an RMS deviation of 0.8 MeV for the 10 nuclei he considered,
encouraging the hope that deviations below 1 MeV might be reached
microscopically.  Of course, open shell nuclei have significant correlation
energies that must be included.  We discuss how this might be done in
Sect. 3 below.  The other problem
is the choice of energy density functional, and there may be reason to
use other forms than the Skyrme or finite-range generalizations.  This
is discussed in the next section.

\section{New forms for the energy functional}

Some perspective on the energy functional can be obtained from the analogous
problem in condensed matter physics.  Correlation effects 
are rather mild in the many-electron problem, and the mean field approach
is very successful.  The energy functional analogous to the Skyrme
is called the local density approximation (LDA), and 
it is widely used to calculate structures of many-atom systems.
Its accuracy for chemical purposes is inadequate, however.
For example, in a comparison of different functionals Perdew \etal 
\cite{pe96} noted  
that the LDA had a mean absolute error of 1.4 eV in a sample of 
small molecules with binding energies in the range 5-20 eV\footnote{
It should be mentioned, however, that the LDA energy functional
is constructed {\it ab initio} without fitting binding data.}.
Two refinements of the energy functional, going beyond the LDA,
make a dramatic improvement in the quality of agreement.  The first
refinement is to include a term in the energy functional that depends
on the gradient of the density.  Gradient terms are already present
in the Skyrme interaction, but in the electron system improvement only 
appears with a nonlinear
functional form of the gradient term,
\be
e(n,\nabla n) \sim  {f(n)\over 1 + a (\nabla n/n)^2}
\label{gradient}
\ee
The mean absolute error decreases by a factor of 4,
to 0.35 eV, when this term is added\footnote{Again, no free parameters
are added with this term; the functional form and dependencies on 
$\nabla n$ and $\tau$ are constructed to simulate the nonlocality of
the exchange.}.   The other improvement is to allow the functional to 
depend on the kinetic energy density $\tau$ and well as on the local
density $n(r)$.  This also is a familiar feature of the Skyrme interaction,
but for the electron case the $\tau$ is combined with the $\nabla n$
dependence in eq. (\ref{gradient}).
The resulting mean absolute error is reduced by almost a factor of
three to give a final mean absolute error of 0.13 eV.  

The nuclear problem is different from the electronic is one important
respect.  In the electronic problem, much of the loss of accuracy is
due to the exchange potential, which is intrinsically nonlocal but
must be treated in a nearly local approximation for computational reasons.  
In contrast, in the
nuclear problem the strong interaction is short-ranged, implying that
the exchange is short-range as well and thus suited to local approximations.
Nevertheless, it might be that more complicated
functional forms such as eq. (\ref{gradient}) 
could be useful in the nuclear problem.
Indeed, this kind of generalization has be examined by Fayans \cite{fa98}.  
He used such terms in an energy functional that was
fit to 100 spherical nuclei.  He obtained an RMS binding error of 1.2 MeV,
a factor of two better than the Gogny or the published Skyrme functionals. 

\section{Correlation energy}

As mentioned earlier, in open shell nuclei the correlation energy associated
with nearly degenerate configurations can be of the order of several
MeV, so the single-configuration mean field approximation is not 
accurate enough for global energy systematics.  

We believe that the following correlations should be considered 
explicitly in the theory:
\hspace*{2cm}$-$center-of-mass delocalization\\
\hspace*{2cm}$-$quadrupole deformations\\
\hspace*{2cm}$-$pairing.\\
These are the obvious correlations associated with symmetry breaking
in the mean field Hamiltonian.
Translational invariance is always broken in finite systems and rotational 
invariance is often broken as well.
Pairing treated by BCS theory violates particle number conservation.

How important are these correlations to the energetics?
For the center of mass, the correlation energy can be estimated in the
harmonic oscillator model (but see below) as $3\hbar \omega_0/4 \approx
30 /A^{1/3}$ MeV.  The resulting magnitude of several MeV is certainly
much larger than the allowable error in a global mass theory.  
However, the energy varies very smoothly and one could question whether it
needs to be treated explicitly as a correlation energy or whether its
effect can be subsumed into the parameters of the mean-field energy 
functional.  This might not be the case because the mean field functional
determines most directly the leading terms in the liquid drop expansion,
varying as $A$ and $A^{-1/3}$.  One should not ask it to simulate a
completely different $A$-dependence.  This argument, and some mean-field
calculations to support, was given recently by Bender \etal \cite{BRRM00}.

The correlation energy associated with the deformations may be thought
of as the energy gained by projecting the states of good angular
momentum out of the deformed intrinsic state which contains many angular
momenta.  Its magnitude can be quite large on our accuracy scale of 
hundreds of keV.  For example, the nucleus
$^{20}$Ne has a Hartree-Fock ground state close in structure to the
[8,0] SU(3) state of the harmonic oscillator.  Combining probabilities of
the different angular states with the energies of the states derived
from the experimental spectrum, one can easily show that the energy 
gained by the projection is of the order of 4 MeV.  A similar number 
was obtained recently in the projected mean-field calculations of $^{24}$Mg 
by ref. \cite{va00}. Thus this correlation energy should be included to 
accuracy
of 10\% or better to achieve the desired accuracy of the mass theory.

The situation with pairing is less severe.
It is certainly necessary to including pairing in a theory of masses,
just to get realistic odd-even mass differences, but the need
to include correlation effects beyond the Hartree-Fock-Bogoliubov theory
is less clear.  With certain simplified pairing interactions the pairing
Hamiltonian can be solved exactly, without making the BCS approximation
\cite{ri64}.  The error in the BCS energy due to the number
nonconservation is of the order of 0.5 MeV \cite{RS80}, Table 11.1. 
This might be significant in the global mass theory and we
shall include it in our discussion.

There are many ways that correlation energies can be calculated.  Most
popular seems to be the obvious method in which the eigenstates
of the symmetry are projected out of the mean-field 
wave function.  If the energy minimization is carried out after the
projection, this method is rather costly to use and probably not
suitable for a global mass theory. For a global theory it
is important that the method be simple computationally and also that
it be systematic, applicable in principle to all possible mean field
solutions.  Particularly important is that it does not introduce
a discontinuity when the mean field solution changes its character.
In the systematic development of many-particle perturbation theory, 
the leading term beyond the mean-field approximation gave the
correlation energy as an integral over the RPA excitation modes.
In a finite system, the RPA correlation formula is given by\cite{RS80,R68}
\begin{equation}
E_{corr}= \frac{1}{2}\left(\sum_i\hbar\omega_i - Tr(A)\right), 
\label{corr}
\end{equation}
where $\omega_i$ is the (positive) 
frequency of the RPA phonon 
and $A$ is the $A$ matrix in the RPA equations. 
This approach was first proposed by Friedrich and Reinhard \cite{fr86}.
It seems to us that this formula is well-suited to the
requirements we need for a systematic mass theory.  
We shall first argue that the formula is adequate in principle, and then
take up the issue of computational feasibility.
In the next section 
we summarize experience with simple
model Hamiltonians that show that eq. (\ref{corr}) 
is more accurate than commonly
used projection techniques, or is easier to calculate.  In the
section following we examine specific algorithms for calculating nuclear
binding energies efficiently.

\section{Experience from simple models}

\subsection{Center-of-mass localization}

The Hamiltonian of two particles interacting through a quadratic
potential $V(r_{12}) = m\omega_0^2 r_{12}^2/2$ is solvable exactly and
also has an analytic mean-field approximate solution.  One might guess
that the RPA correlation energy might give the correction exactly,
because one can derive the RPA by considering quadratic approximations in a 
path integral formulation of the problem. This is indeed the case\cite{HB00}.  
The RPA spectrum has two states, a zero-frequency mode and a finite frequency
mode. Putting these frequencies in eq. (\ref{corr}), 
one finds that the correction
to the mean-field energy $\hbar\omega_0$ is just what is needed to give
the exact energy for the total, $\hbar\omega_0/\sqrt{2}$. 

It is interesting to compare the RPA approach with other ways of
dealing with correlation energies associated with broken symmetries.
In the case of center-of-mass motion, a recipe that is often used is  
to subtract the expectation value of the center of mass 
operator from the mean field energy (e.g. in ref. \cite{BRRM00}). 
With our Hamiltonian, this prescription gives
\begin{equation}
E_{cm}=-\left\langle MF\left|\frac{1}{2}\frac{(p_1 + p_2)^2}{2m}\right|MF
\right\rangle
=-\frac{1}{4}\hbar\omega_0.
\end{equation}
The total $E_{MF} + E_{cm} = 3\hbar\omega_0 / 4 $ is not exact, although
it is close to the exact energy, $\hbar\omega_0/\sqrt{2}$. 
This study shows that for this first kind of correlation the RPA formula 
provides a better method to calculate the associated energy.

\subsection{Deformations}

When the mean field solution is deformed, a continuous symmetry is broken.
As in the above example, a signature of the broken continuous symmetry
is a zero-frequency RPA mode. A model to test theories of the correlation 
energy should thus have a corresponding
continuous symmetry.  We constructed a model with those
properties in ref. \cite{HB00}, making a generalization of the
Lipkin model.  In the original Lipkin model that has been studied
for 40 years, one considers many distinguishable
particles each of which can be in one of two states.  For that
model, the ground-state 
correlation was studied by  Parikh and Rowe \cite{PR68}.  They compared 
various methods of
treating the correlation energy, finding that the RPA formula worked best.
In ref. \cite{HB00}, we 
extended the Lipkin model to a three-state wave function to get sufficient
degrees of freedom for a continuously broken symmetry.  The symmetry is
imposed on the Hamiltonian by requiring it to be invariant under
transformations within two of the three states. 
The two degenerate upper states could be 
thought of as the first excited states of a two-dimensional harmonic 
oscillator, thus allowing deformed wave functions in two dimensions. 
As expected, when the mean field solutions and its RPA excitations are
calculated, ones sees a zero-frequency when the mean field solution is
deformed. There is also another mode at finite frequency.
In this case, the first mode corresponds to rotational motion 
perpendicular to the symmetry axis, while the second mode corresponds 
to the beta vibration.
In the ``spherical" phase, the RPA frequencies for the modes 
are identical and have finite frequency.
Figure 1 shows the RPA frequencies as a function of 
$\chi\equiv V(N-1)/\epsilon$ for the particle number $N=20$, 
where $V$ and $\epsilon$ are the 
strength of the interaction and the single-particle energy, respectively. 
One can clearly see the discontinuity at the critical point $\chi=1$.

Figure 2 compares the ground state energy as a function of $\chi$ 
obtained by several methods. The number of particle $N$ is set to be 20. 
The solid line is the exact solution obtained by numerically 
diagonalizing the Hamiltonian. The dashed line is the ground state energy 
in the Hartree-Fock approximation. 
It considerably deviates from the exact solution through 
the entire range of $\chi$ shown in the figure. The dot-dashed line 
takes into account the RPA correlation energy in addition to the HF 
energy. Clearly the RPA significantly improves the 
results. The corresponding energy as a function of the number of particles
is shown in Fig. 3.  One sees that the RPA correlation energy is 
reliable for large $N$, but may be problematic when there are only
a few valence particles.  That situation is further complicated by
the pairing interaction, which may dominate the mean field solution for
$N=2$.

\subsection{Pairing}

Pairing is the final example of a long-range correlation that significantly
affects the energy.  The mean field approach leads to the BCS theory,
whose ground state has indefinite particle number.  
An early study by
Bang and Krumlinde \cite{BK70} showed that the RPA formula reproduces
the exact correlation energy rather well in a schematic model.  
The RPA method has in fact been used in realistic models of deformed
nuclei\cite{S89}.  
The RPA correlation in the normal phase was studied in ref. \cite{DRS98} 
using the self-consistent version of RPA. 

Kyotoku {\it et al.} 
\cite{KLC96} derived the analytic solution for a model first proposed in ref. 
\cite{hf61}, fermions in a space of
two nondegenerate j-shells interacting with a pairing Hamiltonian.
They were able to solve the model exactly, and then compare the energy
with several approximate methods to calculate the correlation corrections
to the BCS energy.
They found that the ground state energy in 
the BCS + quasi-particle RPA (QRPA) coincides with the exact solution 
at the leading order of an expansion in 1/$N$, the number of particles in
the system.  None of the other methods obtained the correct coefficient
of the leading order contribution.  For example, the well-known method
of Lipkin and Nogami \cite{LN60} gave a result that is only correct
in the limit of a strong pairing force (see Table I in 
ref. \cite{KLC96}). 

In ref. \cite{HB00-pair}, we 
specifically compared 
the RPA with the computationally attractive alternative methods,
testing the behavior across shell closures and considering both 
even- and odd-$N$ systems.  Taking as a test case the two-level
problem with a level degeneracy of $\Omega$ = 8 and a Fermi energy
half way between the levels, the RPA was much superior to the
Lipkin-Nogami method over most of the range of pairing strengths.
This behavior is consistent with 
the result of ref.\cite{KLC96} as we discussed above. 
However, right at the phase transition point
the two methods had comparable errors of opposite sign.

We therefore also looked at a more realistic situation,
varying the particle number 
$N$ rather than the interaction strength $G$. We consider the pairing 
energy in 
oxygen isotopes, taking the neutron 1p and 2s-1d shells as the 
lower and higher levels of the two-level model. 
The pair degeneracy $\Omega$ thus reads $\Omega_1=3$ and $\Omega_2=6$, 
for the lower and the upper levels, respectively. 
The number of particle in a system is given by $N=A-8-2$ for the $^A$O  
nucleus. We assume that the energy difference between the two levels 
$\epsilon$ is given by $\epsilon=41A^{-1/3}$ and the pairing strength 
$G=23/A$. 
The upper panel of Fig. 4 shows the ground state energy as a function of 
$A$. In order to match with the experimental data for the $^{16}$O nucleus, 
we have added a constant $-72.8$ MeV to the Hamiltonian for all the isotopes. 
The exact solutions are denoted by the filled circles. The deviation from 
the BCS approximation (the dashed line) is around 
2 MeV for even $A$ systems and it is around 1.2 MeV for odd $A$ systems. 
This value varies within about 0.5 MeV along the isotopes and shows relatively 
strong $A$ dependence. 
One can notice that the RPA approach (the dot-dashed line) 
reproduces quite well the exact solutions. 
In contrast, the Lipkin-Nogami approach (the 
thin solid line) is much less satisfactory and shows 
a different $A$ dependence from the exact results. 
The pairing gap $\Delta$ in the BCS approximation and in the Lipkin-Nogami 
method is shown separately in the lower panel of Fig. 4. For the Lipkin-Nogami 
method, we show $\Delta+\lambda_2$, which are to be compared with 
experimental data \cite{LN60}. The closed shell 
nucleus $^{16}$O and its neighbor nuclei $^{15,17}$O have a zero pairing 
gap in the BCS approximation, and the Lipkin-Nogami method does not 
work well for these nuclei. 
On the other hand, the RPA approach reproduces the correct $A$ dependence 
of the binding energy. Evidently,  
the RPA formula provides a 
better method to compute correlation energies than the Lipkin-Nogami method, 
especially for shell closures. 

\section{Is RPA computationally feasible?}

We now discuss the practicality of using eq.~(\ref{corr}) to calculate the
correlation energy.  In general, the RPA is computationally more
demanding than the mean field minimization for the ground state
by an order of magnitude or more.  
One must diagonalize an RPA matrix whose dimension is $2D\times 2D$,
where $D$ is the number of particle-hole configurations.  This number
can be 
huge if one is interested in deformed nuclei or heavy spherical nuclei. 
A widely used way around this is to take a residual interaction 
has a separable form. Then the matrix equation to be solved has the
dimension of the rank of the separable interaction; with a single term
it is just the well-known algebraic dispersion relation. 

Given a separable interaction, there are several efficient ways to
the RPA correlation energy (\ref{corr}) without explicitly calculating
all the roots of the dispersion relation.  One method 
was recently proposed by D\"onau {\it et al.}\cite{DAN99} and also by 
Shimizu \etal \cite{SDB99}. 
Instead of directly calculating the RPA correlation energy according 
to eq. (\ref{corr}), one carries out an integration of a function 
which depends on a free response function and its first derivative 
in a complex energy plane. 
An advantage of this method is that one can choose the integration 
path so that the integrand is smooth enough and thus the mesh of the 
numerical integration along such path can be much larger than the actual 
energy intervals of RPA solutions $\omega_{\alpha}$. 
This method is useful particularly when a separable interaction is 
used so that the free response function and its first derivative are 
analytically evaluated. 

Alternatively, one can also use the 
Lanczos-type method proposed in ref. \cite{JBH98} to evaluate the 
RPA correlation energy. As we show below, this 
method quickly converges when the interaction is separable. 
The idea is to define at the outset a 
characteristic operator associated with each kind of correlation.
That operator applied to the mean-field ground state gives an
excited state, which is taken as the the first vector in a space
built up by applications of the Hamiltonian to existing states.
Eq. (\ref{corr}) is then evaluated in the restricted spaces, and the method
would be computational feasible if the convergence is rapid enough.

Suppose $A$ and $B$ matrices in the RPA equation are given by 
\begin{eqnarray}
A_{ij}&=&\epsilon_i\delta_{i,j}+\kappa f_i f_j, \\
B_{ij}&=&\kappa f_i f_j, 
\label{AB}
\end{eqnarray}
where $i$ and $j$ label particle-hole configurations and $f_i$ is normalized 
as $\sum_i f_i^2=1$. 
For such an interaction, the collective operator can be chosen 
as $\psi_i=f_i$. Notice that $f$ is an eigen-vector of the matrix $B$ with 
eigen-value $\kappa$ and also that $B\phi=0$ 
for any vector $\phi$ which is orthogonal to $f$. 
Starting from the initial 
vectors $|X_1\rangle = |\psi\rangle$ and $|Y_1\rangle = 0$, the Lanczos 
manipulation \cite{JBH98} 
transforms the $A$ and $B$ matrices into the form of 
\begin{equation}
\vec{A}'=
\left(\matrix{
e_1 & a_1 & 0 & \cr
a_1 & e_2 & a_2 & \cr
0 & a_2 & e_3 & \cr
  &     &  & \ddots }\right), ~~~~~
\vec{B}'=
\left(\matrix{
\kappa & 0  & 0 & \cr
0 & 0 & 0 & \cr
0 & 0 & 0 & \cr
  &     &  & \ddots }\right). 
\end{equation}
The Lanczos basis for backward scattering amplitude 
$|Y_i\rangle$ remain all zero 
in the manipulation, and thus 
the transformed $B'$ matrix has only one element which is not equal to zero. 
This is not in general the case for non-separable interaction. 
Note that the $B'$ matrix measures the degree of correlation 
in the transformed space. Since it has a very simple form, 
the correlation energy evaluated in the restricted space converges very 
rapidly. 

We have tested the method with RPA matrices given by eq. (\ref{AB}) with 
$\epsilon_i=e_0 + (i-1)\Delta e$. 
The method
works very well when there is a gap in the particle-hole spectrum.
For example, taking a model with $D$=20 particle-hole states and
other parameters $f_i=1$, $\Delta e = e_0/D$, and $\kappa = -0.035 e_0$, the
RPA spectrum has a collective state at $\omega_1 = 0.145 e_0$.  The
correlation energy from eq. (\ref{corr}) with all eigenvalues, is 
$E_{corr} = -0.542 e_0$.  The single-state approximation starting from
the state $\psi_i = f_i/\sqrt{D}$ is only off by 20\%.  The two-state
approximation has an accuracy of 1\%.  The calculational effort to
get this numbers is essentially that of three applications of the
Hamiltonian to the ground-state single-particle wave function, much
less than the effort to get the converged wave functions in the
Hartree-Fock minimization.  See Table I.  

From our point of view, the problem is then to define a reliable separable
interaction that can be used globally in RPA calculations.
For particle-hole residual interactions, there have been proposed a few 
possible ways based on the self-consistency argument to construct a separable 
interaction \cite{BM75,BR97,NKG97,NKGK96,SS81}.   The basic argument
is that the collective motions primarily result in the displacement of
the nuclear surface; thus the transition potential is can be generated
by displacing the self-consistency potential associated with the ground
state density.  In equations, the transition density has a form locally
given by the gradient of the static density, and the transition potential
is the corresponding gradient of the static potential:
\begin{eqnarray}
  \delta \rho &=& a(\Omega) {d \rho_0\over dr } \rightarrow \label{trden} \\
&&  \delta V   = a(\Omega) {d U_0 \over dr}, \label{trpot}
\end{eqnarray}
where $\Omega$ is an angle given the direction to an element of the
nuclear surface.  The separable residual interaction then has the
shape of eq. (\ref{trpot}) and the required magnitude to satisfied 
eq. (\ref{trden}).
Thus
\begin{equation}
v(r_1,r_2) = \delta V (\Omega_1,r_1) \delta V(\Omega_2,r_2)
\left/  \int dr \,\delta V {d \rho\over d r}\right. \,.
\end{equation}

It is amusing to compare the above self-consistent definition with the
microscopic particle-hole interaction proposed by Migdal \cite{M67}.
For the above transition density, his transition potential would be
given by
\begin{equation}
\delta V(r) = V_0 (f^{ex}+(f^{in}-f^{ex})\rho_0(r))\cdot \delta \rho (r).
\end{equation}
The two transition potentials for $^{208}$Pb are compared in Fig. 5. 
We generate the static potential $U_0$ using the velocity-independent part 
of the Skyrme interaction with SIII parametrisation and we use a Fermi 
distribution for a static density $\rho_0(r)$. 
The parameters of the Migdal interaction are given in ref. \cite{RS74}. 
We see both the transition potentials have a similar surface-peaked 
radial dependence. 

The above self-consistency argument can be applied very easily to the
translation mode, where $a(\Omega)=\cos \theta$ for translations in
the $z$-direction.  For the rotational degrees of freedom, we would
use the displacement fields associated with the 5 components of the
quadrupole operator, i.e. $a_{\mu}(\Omega)= Y_{2\mu}(\Omega) ; \mu = -2,-1, 
0,1,2$. Note that the RPA correlations are to be evaluated for spherical as
well as deformed nuclei; a major point of the approach is that the
theory applies to all cases.

It is not so simple to construct a global, separable pairing interaction.
The commonly used forms for the pairing interaction give divergent
results without a cutoff in space of states included in the BCS 
calculation.  
Here we propose a non-local surface-peaked separable form, 
\begin{equation}
v(\vec{r}_1\vec{r}_2,
\vec{r}_1'\vec{r}_2')=\delta(\vec{r}_1-\vec{r}_2)\,
\delta(\vec{r}'_1-\vec{r}_2')\,
\left[-\frac{\chi}{2}\,\frac{dU_0}{dr_1}\,\frac{dU_0}{dr_1'}\,
\sum_{\lambda\mu}Y_{\lambda\mu}(\hat{\vec{r}}_1)
Y^*_{\lambda\mu}(\hat{\vec{r}}_1')\right]. 
\label{pairing}
\end{equation}
The pairing matrix element of this interaction are to be evaluated as
\begin{equation}
\left\langle p_1\bar{p}_1\left|v\right| p_2\bar{p}_2\right\rangle 
=-\frac{\chi}{2}
\sum_{\lambda\mu}
\left\langle p_1\left|\frac{dU_0}{dr}Y_{\lambda\mu}(\hat{\vec{r}})
\right|p_1\right\rangle
\left\langle p_2\left|\frac{dU_0}{dr'}Y^*_{\lambda\mu}(\hat{\vec{r}'})
\right|p_2\right\rangle
\end{equation}
This form is inspired by recent observations that the pairing is essentially 
a surface phenomenon \cite{BBG99,BLSZ99,FS99}. 
A similar surface-peaked separable interaction was used for the particle-hole 
channel in ref. \cite{ABDK96}. 
We have tested the surface-peaked separable pairing interactions 
(\ref{pairing}) by comparing it with the density-dependent delta interaction 
proposed in ref. \cite{BE91}. The problem of the cutoff is much less
severe with this interaction than with a contact interaction, because
the smoothness of $d U_0/dr$ cuts off the radial integrals. 
Our preliminary calculations show that 
the separable pairing interaction can reproduce 
results of the density-dependent pairing interaction reasonably well 
once the strength $\chi$ is adjusted. 
For the purpose of the global binding systematics, the $A$-dependence of 
the strength has yet to be sorted out.   

\section{Summary}

Our goal is to develop a better microscopic theory for the nuclear 
binding systematics. In this paper, we argued that the RPA approach 
provides a promising and computationally tractable way to 
include correlation effects in 
a global model, going beyond the mean-field approximation. 
We have shown that the HF + RPA approach works indeed well using 
simple Hamiltonian models for correlation associated with broken mean-field 
symmetries, namely the center-of-mass localization, 
rotation, and pairing. 
The RPA equation can be easily solved for a separable residual interaction. 
For example, the Lanczos method is quite efficient to solve the RPA 
equation for a separable interaction. 
This contrasts to other popular computational methods such as the generator 
coordinate method and the variation-with-projection method, which 
are rather complicated to apply. 
A work is now in progress to tackle the nuclear masses using 
the HF+RPA approach discussed in this paper. 

\section*{Acknowledgment}
We thank P.-G. Reinhard, W. Nazarewicz, B.A. Brown, and A. Bulgac for 
discussions. 
This work was supported by the U.S. Department of Energy 
under Grant no. DE-FG03-00-ER41132.

\newpage

\begin{table}[hbt]
\caption{Convergence characteristics of the Lanczos algorithm.}
\vspace*{-2pt}
\begin{center}

\begin{tabular}{ccc}

iteration number & $E_{corr}/e_0$ & $\omega_1/e_0$ \\
\hline
1 & $-0.2211 $ & 0.3326 \\
2 & $-0.2688 $ & 0.1539 \\
3 & $-0.2710 $ & 0.1454 \\
\hline
exact & $-0.2711$ & 0.1451 \\
\end{tabular}
\end{center}

\end{table}

\newpage

\begin{figure}
  \begin{center}
    \leavevmode
    \parbox{0.9\textwidth}
           {\psfig{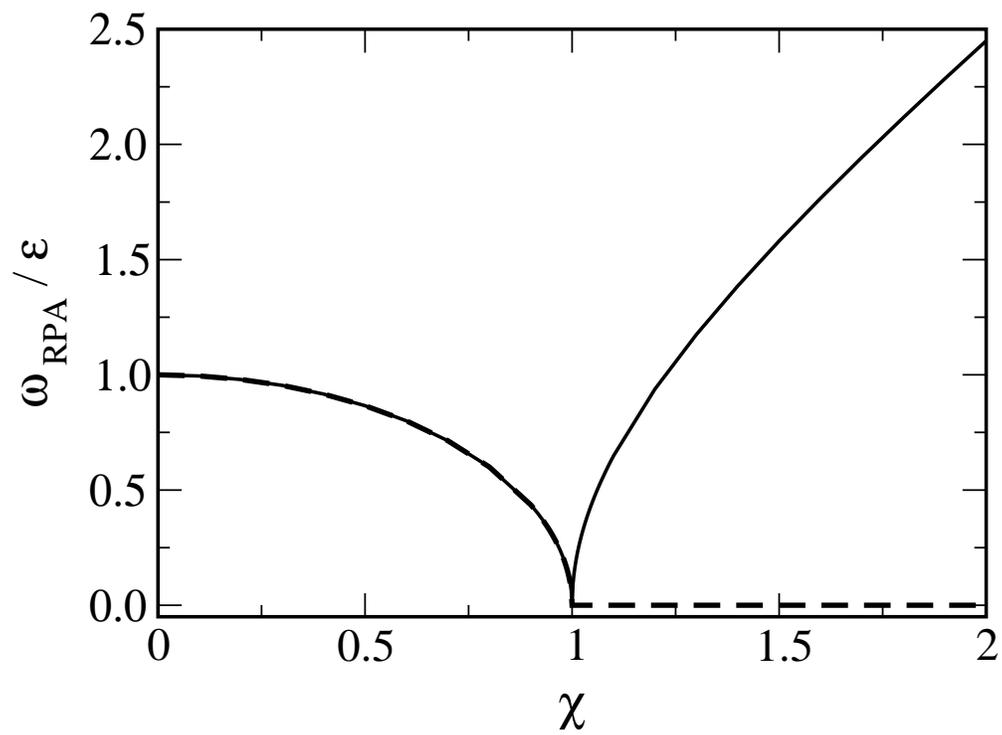}}
  \end{center}
\protect\caption{
RPA frequencies in the three-level Lipkin model as a function of 
$\chi\equiv V(N-1)/\epsilon$. The number of particle $N$ is 
chosen to be 20. 
}
\end{figure}

\newpage

\begin{figure}
  \begin{center}
    \leavevmode
    \parbox{0.9\textwidth}
           {\psfig{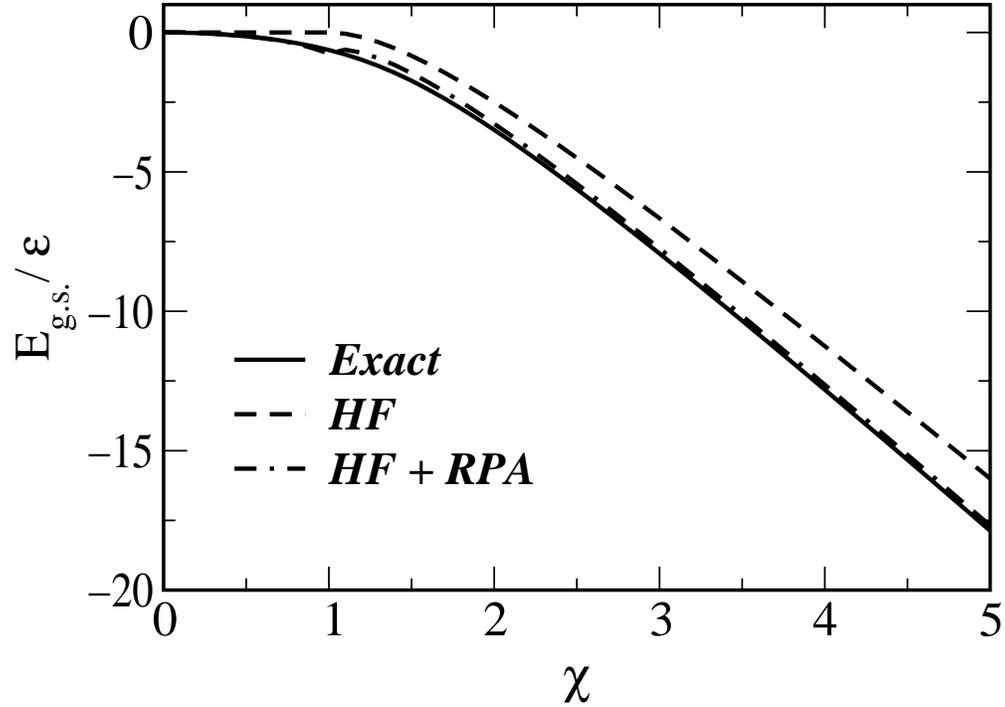}}
  \end{center}
\protect\caption{
Comparison of the ground state energy of the three-level Lipkin model 
obtained by several methods. 
The solid line is the exact numerical solution. The ground state energy 
in the Hartree-Fock approximation is denoted by the dashed line, while 
the dot-dashed line takes into account the RPA correlation energy in 
addition to that. 
}
\end{figure}

\newpage

\begin{figure}
  \begin{center}
    \leavevmode
    \parbox{0.9\textwidth}
           {\psfig{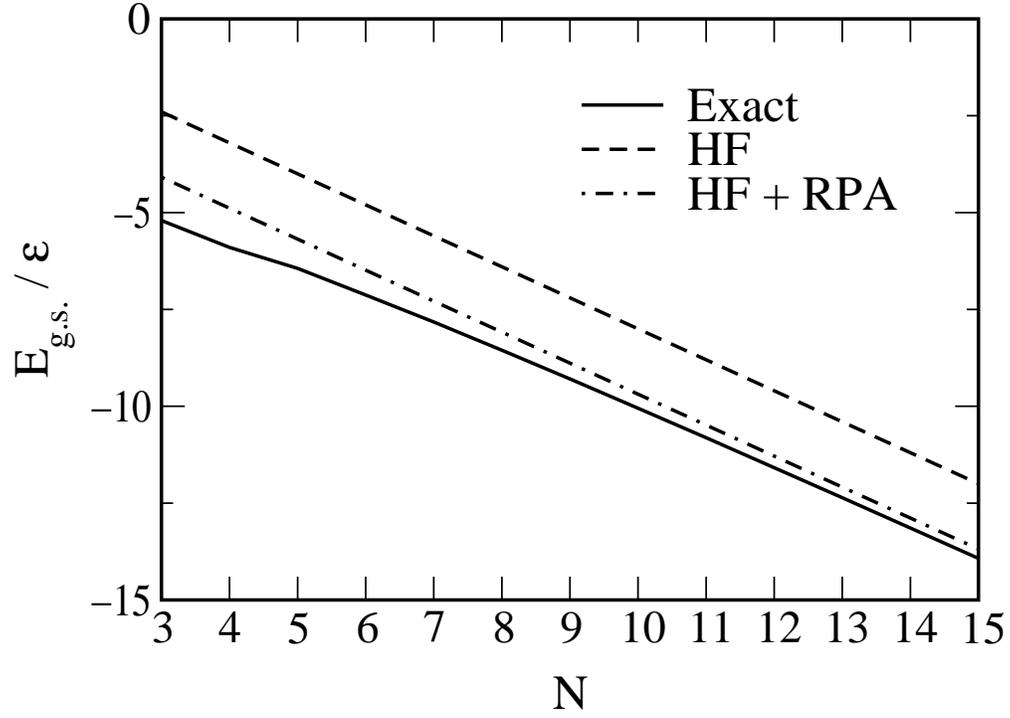}}
  \end{center}
\protect\caption{
The ground state energy of the three-Level Lipkin model 
as a function of $N$ for $\chi$=5.0. 
}
\end{figure}

\newpage

\begin{figure}
  \begin{center}
    \leavevmode
    \parbox{0.9\textwidth}
           {\psfig{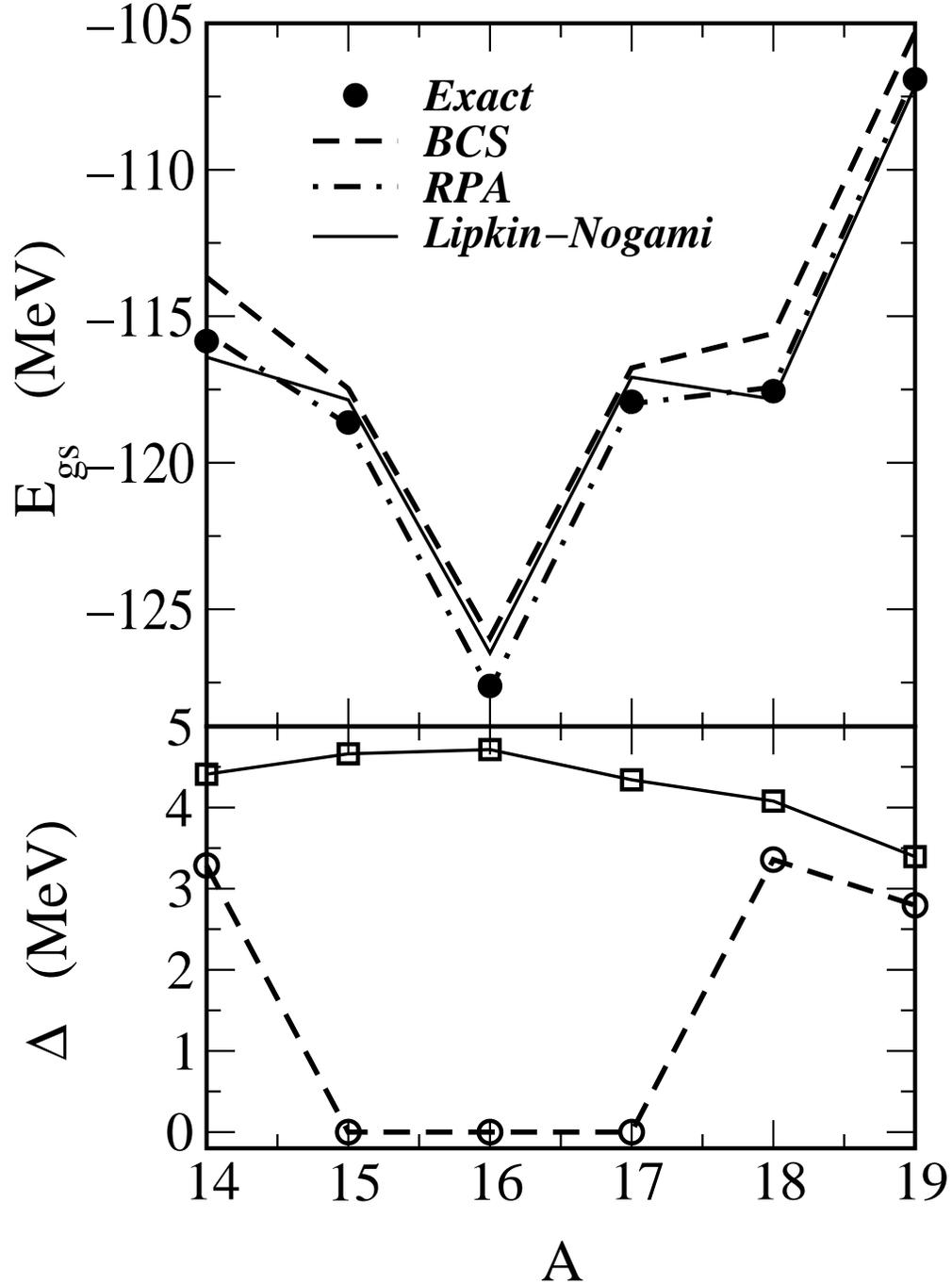}}
  \end{center}
\protect\caption{
The ground state energy $E_{gs}$ (the upper panel) and the 
pairing gap (the lower panel) for oxygen isotopes 
estimated with the two-level model 
as a function of the mass number.  
The exact results are denoted by the filled circles, while 
the meaning of each line is the same as in Fig. 2. 
For the pairing gap in the Lipkin-Nogami method, $\lambda_2$ is added to 
the pairing gap $\Delta$. 
}
\end{figure}

\newpage

\begin{figure}
  \begin{center}
    \leavevmode
    \parbox{0.9\textwidth}
           {\psfig{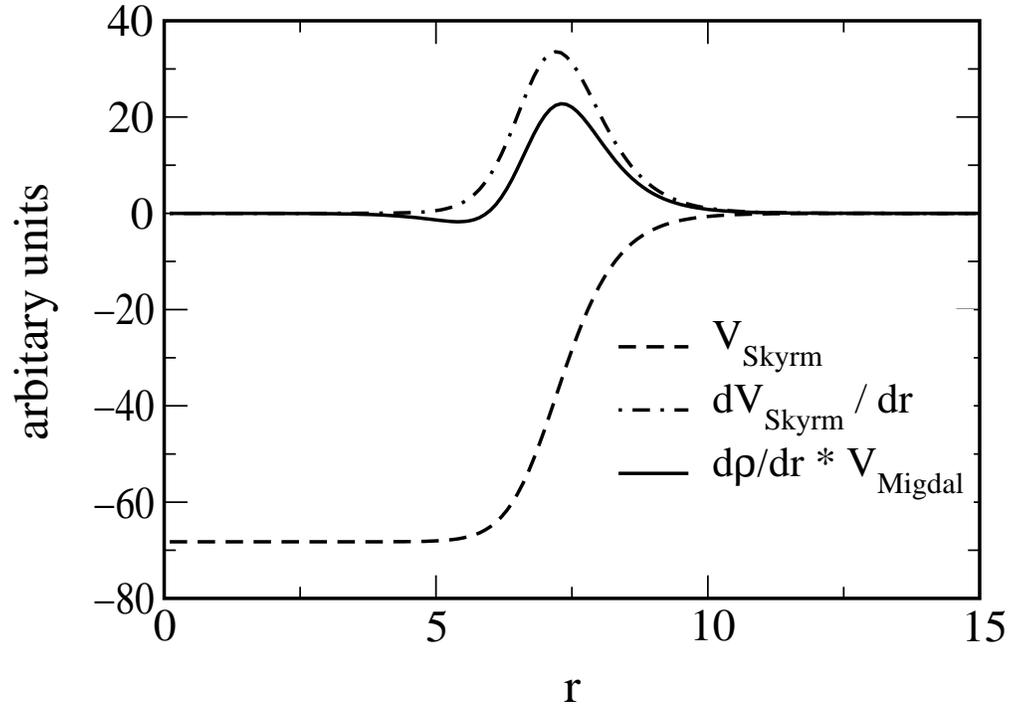}}
  \end{center}
\protect\caption{
Comparison between transition potentials obtained with the Skyrm and the 
Migdal interactions.}
\end{figure}

\end{document}